\documentclass[twocolumn,showpacs,pra,nofootinbib]{revtex4}
\usepackage{amsmath}
\usepackage{latexsym}
\usepackage{epsfig}

\newcommand{\clP}{{\cal P}}

\newcommand{\clF}{{\cal F}}
\newcommand{\clI}{{\cal I}}
\newcommand{\hh}{\tilde{h}}

\newcommand{\rgl}{\rangle}
\newcommand{\lgl}{\langle}

\newcommand{\be}{\begin{equation}}
\newcommand{\ee}{\end{equation}}
\newcommand{\bea}{\begin{eqnarray}}
\newcommand{\eea}{\end{eqnarray}}

\begin{document}

\title{Hyperdiffusion of quantum waves in random photonic lattices }
\author{Alexander Iomin }
\affiliation{Department of Physics, Technion, Haifa, 32000, Israel \\
\\
Phys. Rev. E \textbf{92}, 022139 (2015)}

\begin{abstract}
A  quantum-mechanical analysis of hyper-fast (faster than ballistic) diffusion of a quantum wave packet in random optical lattices is presented. The main motivation of the presented analysis is experimental demonstrations of hyper-diffusive spreading of a wave packet in random photonic lattices [L. Levi \textit{et al.}, Nature Phys. \textbf{8}, 912 (2012)].
A rigorous quantum-mechanical calculation of the mean probability amplitude is suggested, and it is shown that the power law spreading of the mean squared displacement (MSD) is $\lgl x^2(t)\rgl\sim t^{\alpha}$, where  $2<\alpha\leq 3$. The values of the transport exponent $\alpha$ depend on the correlation properties of the random potential $V(x,t)$, which describes random inhomogeneities of the medium.
In particular, when the random potential is $\delta$ correlated in time, the quantum wave packet spreads according Richardson turbulent diffusion with the MSD $\sim
t^3$. Hyper-diffusion with $\alpha=12/5$ is also obtained for arbitrary correlation properties of the random potential.

\end{abstract}

\maketitle

\section{Introduction}
Recently, it has been demonstrated, experimentally and numerically \cite{segev3}
that space-time disordered media accelerate the transport in a way, when an initial
wave packet spreads at a rate faster than ballistic with the mean squared
displacement (MSD) $\langle{x}^2(t)\rangle\sim t^{12/5}$.
This effect has been explained in the framework of classical mechanical
approach due to continuous expansion of the transverse momentum
spectrum in an arbitrary space-time random potential
\cite{segev3,wilkinson,fishman}. In this paper we suggest a quantum-mechanical
explanation of this experimental observation of the disorder enhanced transport in
photonic lattices \cite{segev3}, which is a more general approach for
a quantum wave packet spreading in randomly inhomogeneous media
\cite{segev3,segev1,segev2}.

An investigation of wave spreading in randomly inhomogeneous media is a long
lasting problem, which has been well reviewed already more than thirty years ago
\cite{dashen,kliatskin}, where  a variety of applications have been considered,
and this theory has also a strong impact on statistical methods in physics \cite{zj}
(see also recent review \cite{UPN2004}).

The main objective of the present research is an estimation of the mean squared displacement
(MSD) of the wave packet spreading in the transversal direction (which is the $x$ axis)
under its propagation along a wave-guide. Here the main accent is made on the rigorous calculation of the mean probability amplitude.
It is known that a wave propagation with the wavelength $\lambda$ in a long
range--dependent wave-guide can be described by the parabolic equation in
the limit of a small-angle propagation \cite{kliatskin,tappert}.
This equation  corresponds formally  to the Schr\"odinger equation with an effective
Planck constant of the order of $\lambda$ .
Formally, the longitudinal coordinate plays a role of an effective time $t$,
and the dynamics takes place in a random potential $V(x,t)$, which is a space-time dependent noise. A rigorous quantum-mechanical consideration
is suggested for this Langevin-Schr\"odinger equation, and the wave function is obtained
as functional of $V(x,t)$. We show that the quantum process of spreading depends on
the time correlation properties of the random potential.
We obtain the hyperfast spreading of the quantum wave packet with the MSD
$\langle x^2(t) \rangle\sim t^{\alpha}$ with the transport exponent $2<\alpha\le 3$.
When the noise is a Markov ($\delta$-correlated)
process, the quantum wave packet dynamics corresponds to Richardson diffusion \cite{richardson}
with the MSD of the order of $t^3$. This classical turbulent diffusion is obtained
here by the rigorous quantum mechanical treatment.
A phenomenological statistical
approach dated back to works by Kolmogorov and Obukhov \cite{frisch,monin_yaglom}
suggested this turbulent acceleration by means of a Gaussian $\delta$-correlated
noise \cite{obukhov},
added to the dynamical system $\ddot{x}+V(t)=0$. In this case, due to the noise term
$V(t)$,  Richardson diffusion takes place with the MSD $\langle{x}^2(t)\rangle\sim t^3$,
which is due to the diffusive spread of the velocity profile $\langle\dot{x}^2(t)\rangle\sim t$.
In quantum mechanics, the Kolmogorov-Obukhov approach was first applied
in Ref.~\cite{prl48} to study a non-diffusive motion, where a Gaussian $\delta$ correlated in time random process was treated in the framework of the Furutsu-Novikov formula \cite{furutsu,novikov}  for the mean probability amplitudes \cite{kliatskin,UPN2004}. Recently, it was applied
to described a tracer behavior \cite{barkai} for an explanation of a limiting case of
experimental realization of quantum-mechanical superdiffusion of ultra cold atoms  \cite{nir}.

However, in real experimental realizations of  the disorder in
photonic lattices \cite{segev1,segev2}, the random potential does not possess this
Markov property due to the finite size of the optical wavelength $\lambda$.
Therefore, the quantum dynamics is considered in an arbitrary random
potential $V(x,t)$, which is correlated in both time and space.
In this case a rigorous quantum-mechanical analysis
cannot be performed, and a suitable approximation is suggested to treat
this random quantum dynamics.

It is well known that the quantum dynamics
can be described by a complex Gaussian kernel\footnote{This presentation of the quantum dynamics
by means of an auxiliary Markov field in the framework of the Feynman-Kac formula does not suppose any Markovian property of quantum mechanics.} in
functional integration \cite{feynman,kac}. When $V(x,t)$ is $\delta$ correlated in time,
it does not affect the quantum Gaussian paths in the functional integration that
makes it possible to treat the quantum mechanics rigorously, like
in the Richardson diffusion case.
The situation changes essentially, when the random
process is strongly correlated. Then the quantum paths are affected by the inhomogeneities
of the media. The rigorous analysis is impossible, and terms, which are responsible for
this ``intertwine'', are treated approximately by averaging this part of the quantum paths.
The suggested averaging procedure is performed in self-consistency with the quantum spreading,
and as a result of this, we obtain hyperdiffusion of the quantum packet spreading,
when the MSD is of the order of $t^{12/5}$.  which coincides with a result obtained in
Refs.~ \cite{wilkinson,fishman} in the ray dynamics limit.

Therefore, as the  result of the parabolic equation approximation of the wave process,
the  wave spreading
in randomly inhomogeneous media is investigated in the framework of quantum mechanics with a random potential, which  is the Langevin Schr\"odinger equation.
An important motivation for this analysis is experimental
investigations of
quantum wave packet spreading in random optical lattices \cite{segev3,segev1,segev2}.
Another interesting motivation of the present analysis is investigation
(experimental and theoretical) of sound waves spreading in underwater acoustics
in the presence of random environments (see e.g. recent results in Refs. \cite{UFN,tomsovic}).

\subsection{Parabolic equation approximation}
The method of parabolic equation approximation
was first applied by Leontovich in studying radio-waves spreading \cite{leontovich}
and later it has been developed in detail by Khohlov \cite{khohlov} (see also \cite{tappert}).
Parabolic equation for monochromatic light propagation in two
dimensional randomly inhomogeneous media reads \cite{kliatskin,segev1,segev3}
\begin{equation}\label{paraxial_eq}
i\partial_z\Psi=[-\frac{1}{2k}\partial_x^2\Psi-\frac{k}{n_0}\Delta n(z,x)]\Psi\, .
\end{equation}
Here $\Delta n(x,z)$ is local fluctuations of refractive
index $n=n_0+\Delta n$, and $z$ is the propagation direction
of the wave with the wave index $k=2\pi n_0/\lambda$, therefore
an effective semiclassical parameter is of the order of
$1/k$.
In what follows it is convenient to work with dimensionless
variables and parameters. Taking into account that Eq.
(\ref{paraxial_eq}) has a form of a Schr\"odinger equation,
one defines the dimensionless effective time $t=z/\lambda$ and
the dimensionless effective Planck constant
$\hh=\frac{1}{k\lambda}=\frac{1}{2\pi n_0}$,
then, the dimensionless quantum momentum is
$\frac{\lambda}{k\lambda}\partial_x=\hh\partial_x$, where
$x/\lambda\rightarrow x$.
Note that the wavelength in the experimental
setup is $\lambda\sim 0.514{\rm \mu m}$ and
$\Delta n/n_0\sim 10^{-4}\ll 1$ \cite{segev3}.
Therefore, the effective Planck constant is
a small semiclassical parameter.

\section{Quantum Langevin equation}
Formally, the wave function $\Psi(x,t)$ describes the dynamics of a
quantum wave packet (particle) in random time-dependent optical potential $V(x,t)$, and it is governed
by the Schr\"odinger equation, which reads
\begin{equation}\label{Schrodinger}
\partial_t\Psi(x,t)=[i\hh\partial_x^2/2+iV(x,t)/\hh]\Psi(x,t)\,
\end{equation}
with the initial condition
\begin{equation}\label{initial}
\Psi(x,t=0)=\Psi_0(x)\, .
\end{equation}
Considering the optical random potential $V(x,t)$ as an expansion of a
quasiperiodic function, one has \cite{segev3,fishman}
\begin{equation}\label{potentialV}
V(x,t)=\frac{1}{\sqrt{N}}\sum_{m=-N}^NA_m\exp(ik_mx-i\omega_mt)+c.c.\, ,
\end{equation}
where the coefficients of the expansion $A_m$ are random complex values, while $k_m$ and $\omega_m$ are independent random real values.
Denoting averaging over the  Gaussian ensemble by $\lgl\dots\rgl_{V}$, we obtain
that $A_m$ are
controlled by a Gaussian distribution with the averaging property
\begin{equation}\label{EAm}
\lgl A_m\rgl_{V}=\lgl A_mA_n\rgl_{V}=0\, , ~~~\lgl A_m^*A_n\rgl_{V}=\sigma^2\delta_{m,n}\, .
\end{equation}
From this property, one obtains for the 1D space-time dependent potential $V(x,t)$
\begin{equation}\label{EV}
\lgl|V(x.t)|^2\rgl_{V}=2\sigma^2\, .
\end{equation}
Note that this formulation of the random potential is general and corresponds to the
experimental setup \cite{segev3}.
Since $V(x,t)$ is a random function, the Schr\"odinger Eq. (\ref{Schrodinger}) is a
Langevin equation with a multiplicative noise potential $V(x,t)$.

Following Ref. \cite{kliatskin}, this equation can be solve exactly.
The solution of Eq. (\ref{Schrodinger}) can be presented in the form of a
functional integration over an
auxiliary Gaussian field $\lambda(t)$. The details of the calculation can be found in Ref.~\cite{UPN2004}.
However, here we present an alternating way of the solution, which is more suitable for the quantum-mechanical consideration.

\subsection{Solution of Langevin equation}
A formal integration of the Schr\"odinger Eq. (\ref{Schrodinger}) yields a $T$
ordered (time ordered) form of the evolution operator, which acts on the initial wave function
\begin{equation}\label{Tord_Psi}
\Psi(x,t)=\hat{T}\exp\Big[\frac{i\hh}{2}\int_0^t\partial_x^2d\tau+\frac{i}{\hh}
\int_0^tV(x,\tau)d\tau\Big]\, .
\end{equation}
Under the sign of the time ordering operator $\hat{T}$, all values are commuted, and
the kinetic and potential exponentials can stay separate.
Therefore, for the kinetic term, one applies the Hubbard-Sratonovich
transformation \cite{Strat,Hab}
\begin{eqnarray}\label{QuantHS}
&{}&\exp\Big[\frac{i\hh}{2}\int_0^t\partial_x^2d\tau\Big]=
\int\prod_{\tau}\frac{d\lambda(\tau)}{\sqrt{2\pi\hh i}} \nonumber \\
&\times & \exp\Big[\frac{i}{2\hh}\int_0^td\tau\lambda^2(\tau)\Big]\cdot
\exp\Big[\int_0^td\tau\lambda(\tau)\partial_x\Big] \, .
\end{eqnarray}
Taking into account that the last exponential acts as a shift operator, one obtains the solution
\begin{eqnarray}\label{sol_Psi}
&{}&\Psi(x,t)=\int\prod_{\tau}\frac{d\lambda(\tau)}{\sqrt{2\pi\hh i}}
\exp\left[\frac{i}{2\hh}\int_0^td\tau\lambda^2(\tau)\right] \nonumber \\
&\times &\Psi_0\Big(x+\int_0^td\tau\lambda\Big)\exp\left[\frac{i}{\hh}\int_0^td\tau
V\Big(x+\int_{\tau}^t d\tau'\lambda\, ,\tau\Big)\right]\, . \nonumber \\
&{}&
\end{eqnarray}
Therefore, the quantum-mechanical estimation of the MSD $\langle x^2(t)\rangle$ leads
to two standard procedures of averaging. First one obtains a mean probability amplitude
\footnote{Note that an important information about the random process is carried by
the correlation function of $V(x,t)$, which can be obtained by the ensemble averaging.}
$\lgl|\Psi(x,t)|^2\rgl_{V}$ by averaging of the obtained result in Eq. (\ref{sol_Psi})
over all realizations of the random field $V(x,t)$, and then performs a standard 
quantum-mechanical calculation of the MSD. Therefore the MSD reads
\begin{equation}\label{QMSD}
\langle x^2(t)\rangle=\int x^2 \lgl|\Psi(x,t|^2\rgl_{V}dx\, .
\end{equation}

\section{Mean probability amplitude}

For the random quantum process, the physical characteristics are described by
the mean probability amplitude  (MPA), or distribution function $\rho(x,t)$, obtained from the random wave function (\ref{sol_Psi}) by averaging over the Gaussian distribution
\begin{equation}\label{rho}
\rho(x,t)=\lgl|\Psi(x,t)|^2\rgl_{V}\, .
\end{equation}
Obviously, this value is normalized $\int dx\rho(x,t)=1$.
Following \cite{sokolov,bi1987,pre70_2004}, let us obtain this normalization condition.
The initial condition can be presented by means of the Fourier integration
\begin{equation}\label{FourierPsi}
\Psi_0(x)=\frac{1}{2\pi}\int_{-\infty}^{\infty}\bar{\Psi}_0(k)e^{-ipx}dp\, .
\end{equation}
Substituting this expression in Eq. (\ref{sol_Psi}), one obtains for the MPA
\begin{eqnarray}\label{MPDF}
\rho(x,t)&=&\int\prod_{\tau}\frac{d\lambda_1(\tau)d\lambda_2(\tau)}{2\pi\hh}
\exp\left[\frac{i}{2\hh}\int_0^t(\lambda_1^2-\lambda_2^2)d\tau\right] \nonumber \\
&\times &\int\frac{dp_1dp_2}{4\pi^2}\bar{\Psi}_0(p_1)\bar{\Psi}_0^*(p_2)
\exp[-ix(p_1-p_2)] \nonumber \\
&\times &\exp\left\{-i\int_0^t[p_1\lambda_1(\tau)-p_2\lambda_2(\tau)]d\tau\right\}
\nonumber \\
&\times &\Big\lgl\exp\Big[\frac{i}{\hh}\int_0^t\Big(
V(\tilde{x}_1\, , \tau)-V^*(\tilde{x}_2 \, ,\tau)\Big)d\tau\Big]\Big\rgl_{V}\, ,  \nonumber \\
&{}&
\end{eqnarray}
where $\tilde{x}_j=x+\int_{\tau}^t\lambda_j(\tau')d\tau'$ and $j=1,2$.

\subsection{Integration over the Gaussian distribution}

Now one can treat the random potential term by integration over the $2N+1$ dimensional
Gaussian packet, and this procedure coincides with integration over
many-dimensional coherent states \cite{karruthers_nieto}
\begin{equation}\label{CohStates}
d\Big[\clP\Big(\{A_m^*\,,A_m\}\Big)\Big]=\exp\Big(-\sum_m|A_m|^2/\sigma^2\Big)
\prod_m\frac{d^2A_m}{\pi\sigma^2}\, ,
\end{equation}
where $d^2A_m=d[Re(A_m)]d[Im(A_m)]$. Therefore, after taking into account Eq. (\ref{potentialV}), the ensemble averaging procedure corresponds to the following $2N+1$ dimensional
integration
\begin{eqnarray}\label{averaging}
\lgl\dots\rgl_{V}&=&\prod_m\int\frac{d^2A_m}{\pi\sigma^2}\exp\Big[-\sum|A_m|^2/\sigma^2\Big]
\nonumber \\
&\times &\exp\Big[\sum_m\Big(A_m\alpha_m^*-A_m^*\alpha_m\Big)\Big]\, ,
\end{eqnarray}
where $\alpha_m$ is the following complex function
\begin{equation}\label{alpha}
\alpha_m=\frac{i}{\hh\sqrt{N}}\int_0^td\tau\Big[e^{-ik_m\tilde{x}_1(\tau)+i\omega_m\tau}
-e^{-ik_m\tilde{x}_2(\tau)+i\omega_m\tau}\Big]\, .
\end{equation}
Using the property of integration of coherent states \cite{karruthers_nieto}, namely
\begin{equation}\label{car_n_9_7}
\int\frac{d^2\beta}{\pi}e^{-|\beta|^2}e^{\alpha^*\beta}f(\beta^*)=f(\alpha^*)\, ,
\end{equation}
one obtains from the integration in Eq. (\ref{averaging})
\begin{equation}\label{random_term}
\lgl\dots\rgl_{V}=
\exp[-\sigma^2\sum_m|\alpha_m|^2]\equiv\clF\Big[\lambda_1(\tau),\lambda_2(\tau)\Big] \, .
\end{equation}
The next step of the quantum analysis is functional
integration over the auxiliary Gaussian fields
$\lambda_1$ and $\lambda_2$.
However, the exact quantum-mechanical treatment
is possible only for the $\delta$ correlated in time random potential $V(x,t)$
\begin{equation}\label{delta_correlator}
\lgl V^*(x,t)V(x',t')\rgl_{V}=C(x,t;x',t')=C(x,x')\delta(t-t')\, ,
\end{equation}
where $C(x,x)=2\sigma^2$ (\textit{cf}. Eq. (\ref{EV})).
First, we consider this case, noting that the restriction of $\delta$ correlation corresponds also to the Obukhov mechanism of Richardson diffusion \cite{obukhov}.

\section{Richardson diffusion}

Richardson diffusion \cite{richardson} was the first
phenomenological observation of developed turbulence \cite{falkovich}, and
this phenomenon has been discussed in a variety of experimental and
numerical studies, see reviews \cite{falkovich,frisch} and as admitted in
\cite{falkovich,baule}, it still lacks sufficient experimental confidence.

Let us define the property of $V(x,t)$ by means of the spectral density
$S(k,\omega)$ of the correlation function $C(x,t;x',t')$ with the $\delta$ correlated constraint
(\ref{delta_correlator}). Following Refs. \cite{wilkinson,fishman}, we present the correlation function in the following translational invariant in space and time form
\begin{eqnarray}\label{CorrelatorC}
&{}&C(x,x')\delta(t-t')=
\frac{\sigma^2}{N}\sum_m\Big[e^{ik_m(x-x')-i\omega_m(t-t')}+c.c.\Big]
\nonumber \\
&=&\sigma^2\int dk\int d\omega \tilde{S}(k,\omega)
\Big[e^{i[k (x - x')-\omega(t-t')]} + c.c.\Big] \nonumber \\
&=&\sigma^2\int dk S(k)\cos[k(x-x')]\delta(t-t') \, ,
\end{eqnarray}
where $S(k)=4\pi\tilde{S}(k,\omega)$.

Using this delta correlated property, one can describe the dynamics of
$|\alpha_m(t)|^2$ in Eq.  (\ref{random_term}) by means  of the spectral
density $S(k)$. Substituting Eq. (\ref{alpha}) in Eq. (\ref{random_term}) and taking into account Eq. (\ref{CorrelatorC}), one obtains
\begin{eqnarray}\label{random_SpecD}
&{}&\clF[\lambda_1(\tau),\lambda_2(\tau)]=\exp\Big[
-\sigma^2\sum_m|\alpha_m|^2\Big] \nonumber \\
&=&\exp\Big\{-\frac{\sigma^2}{\hh^2N}
\int_0^td\tau_1\int_0^td\tau_2\sum_{m=-N}^N  \nonumber \\
&\times&\Big[e^{ik_m\tilde{x}_1(\tau_1)-i\omega_m\tau_1}
-e^{ik_m\tilde{x}_2(\tau_1)-i\omega_m\tau_1}\Big]\nonumber \\
&\times&\Big[e^{-ik_m\tilde{x}_1(\tau_2)+i\omega_m\tau_2}
-e^{-ik_m\tilde{x}_2(\tau_2)+i\omega_m\tau_2}\Big]\Big\}\nonumber \\
&=&\exp\left\{-\frac{\sigma^2}{\hh^2}\int_0^td\tau\int_{-\infty}^{\infty}dkS(k)\right.
\nonumber \\
&\times&\left.\Big\{1-\cos\Big[k
\int_{\tau}^t\Big(\lambda_1(\tau')-\lambda_2(\tau')\Big)d\tau'\Big]\Big\}\right\}\, .
\end{eqnarray}
To take the  functional integrals over the auxiliary fields $\lambda_1(\tau)$ and $\lambda_2(\tau)$, one performs the following linear change of the fields \cite{sokolov}
\begin{eqnarray}\label{lambda-mu-nu}
\lambda_1(\tau)&=&2\mu(\tau)+\hh\nu(\tau)/2 \nonumber \\
\lambda_2(\tau)&=&2\mu(\tau)-\hh\nu(\tau)/2\, ,
\end{eqnarray}
where the Jakobian of the transformation is $\hh$
for each value of $\tau$.
Then the functional part of the integrand in Eq. (\ref{MPDF}) reads
\begin{eqnarray}\label{mu_nu}
&{}&\prod_{\tau}\frac{d\mu(\tau)d\nu(\tau)}{2\pi}
\exp\Big[i\int_0^t\mu(\tau)\nu(\tau)d\tau\Big] \nonumber \\
&\times& \exp\Big[-2i(p_1-p_2)\int_0^t\mu(\tau)d\tau \nonumber \\
&-&\frac{i\hh}{4}(p_1+p_2)\int_0^t\nu(\tau)d\tau\Big]
\clF[\nu(\tau)]\, ,
\end{eqnarray}
where we use the fact that
$\clF[\lambda_1(\tau),\lambda_2(\tau)]=\clF[\lambda_1(\tau)-\lambda_2(\tau)]$, which
follows from Eq. (\ref{random_SpecD}).
Taking integration over $x$ in Eq. (\ref{MPDF}) one obtains the $\delta$ function
$\delta(p_1-p_2)$. Then, functional integration over $\mu(\tau)$ yields the delta functions
$\prod_{\tau}\delta(\nu(\tau))$, since the rest of the integrand does not depend on $\mu(\tau)$.
Finally, after integration over $\nu(\tau)$ one obtains that the MPA is normalized to 1
\begin{equation}\label{normalization}
\int_{-\infty}^{\infty}\rho(x,t)dx=1\, .
\end{equation}

\subsection{Mean squared displacement}
Handling the exact expression of the MPA, we arrive at the main objective of the work
and can evaluate the rate of the wave packet spreading by calculation of the MSD
$\langle x^2(t)\rangle$  in the (transversal) $x$ direction.
Taking into account Eqs. (\ref{MPDF}) and (\ref{random_SpecD}), one obtains
for the MSD
\begin{eqnarray}\label{MSD_1}
&{}&\langle x^2(t)\rangle=\int_{-\infty}^{\infty}\rho(x,t) x^2dx \nonumber \\
&=&\int\frac{dp_1dp_2}{4\pi^2}\bar{\Psi}_0(p_1)\bar{\Psi}_0^*(p_2)\delta^{(2)}(p_1-p_2)
\nonumber \\
&\times &\int\prod_{\tau}\frac{d\mu(\tau)d\nu(\tau)}{2\pi}
\exp\Big[i\int_0^t\mu(\tau)\nu(\tau)d\tau\Big] \nonumber \\
&\times & \exp\Big[-2i(p_1-p_2)\int_0^t\mu(\tau)d\tau \nonumber \\
&-&\frac{i\hh}{4}(p_1+p_2)\int_0^t\nu(\tau)d\tau\Big]
\clF[\nu(\tau)]\, ,
\end{eqnarray}
where we use the following definition of the second derivative of the delta function
$\delta^{(2)}(p_1-p_2)\equiv\partial_{p_1}\partial_{p_2}\delta(p_1-p_2)$.
Now, we can repeat the previous calculations of Eqs. (\ref{MPDF}), (\ref{random_SpecD}),
and (\ref{mu_nu}). Functional integration over $\mu(\tau)$ yields
$\prod_{\tau}\delta[\nu(\tau)-2(p_1-p_2)]$.  Therefore, functional integration over $\nu(\tau)$
is rigorous, as well. Performing integration with $\delta^{(2)}(p_1-p_2)$, one obtains finally
for the MSD
\begin{equation}\label{MSD_2}
\langle x^2(t)\rangle= P_0^2 t^2+\frac{D_0}{3}t^3\, .
\end{equation}
Here the first term $\sim t^2$ describes a well known wave packet spreading in
homogeneous media with the mean squared momentum $$ P_0^2=\frac{\hh^2}{2\pi}
\int_{-\infty}^{\infty}p^2|\hat{\Psi}_0(p)|^2dp\, .$$
The second term, which is obtained by the rigorous quantum
mechanical calculations,  is of a pure classical nature and corresponds to
Richardson diffusion \cite{richardson}. However, its contribution in the quantum process
of the wave packet spreading is dominant $\sim D_0t^3$, where the generalized diffusion coefficient is
\begin{equation}\label{D0}
D_0=\sigma^2\int_{-\infty}^{\infty}k^2S(k)dk\,.
\end{equation}

\section{Hyper-diffusion}
It should be stressed that the experimental realization of photonic lattices
with the $\delta$-correlated random potential is technically impossible
\cite{segev3,segev1,segev2}. Therefore, the estimation
of the MSD for the realistic arbitrary correlated random potential $V(x,t)$ leads to
essential complication of the analysis. Let us return to Eqs. (\ref{CorrelatorC}) and
(\ref{random_SpecD}) in a general form of the spectral density $S(k,\omega)$.
The correlation function reads
\begin{eqnarray}\label{CorrelatorC_Hyper}
&{}&C(x-x';t-t')=
\frac{\sigma^2}{N}\sum_m\Big[e^{ik_m(x-x')-i\omega_m(t-t')}+c.c.\Big]
\nonumber \\
&=&\sigma^2\int dk\int d\omega \tilde{S}(k,\omega)
\Big[e^{i[k (x - x')-\omega(t-t')]} + c.c.\Big] \, . \nonumber \\
\end{eqnarray}
In this case, the functional action $\clI[\lambda_1(\tau),\lambda_2(\tau)]=-\sigma^2\sum_m|\alpha_m|^2$
in $$\clF[\lambda_1(\tau),\lambda_2(\tau)]=\exp\{\clI[\lambda_1(\tau),\lambda_2(\tau)]\}$$
in Eq. (\ref{random_SpecD}) is a more complicated expression, which is not treatable rigorously.
After some algebraic manipulations, this reads
\begin{eqnarray}\label{action_clI}
&{}&\clI[\lambda_1(\tau),\lambda_2(\tau)]=
-\frac{\sigma^2}{\hh^2}\int_0^td\tau_1\int_0^td\tau_2\int_{-\infty}^{\infty}dk \int_{-\infty}^{\infty}d\omega
\nonumber \\
&\times & S(k,\omega)\Big[e^{ik\int_{\tau_2}^{\tau_1}\lambda_1d\tau'- \omega(\tau_1-\tau_2)}
\Big(1-e^{ik\int_{\tau_1}^t(\lambda_1-\lambda_2)d\tau'}\Big) \nonumber \\
&+& e^{ik\int_{\tau_2}^{\tau_1}\lambda_2d\tau'- \omega(\tau_1-\tau_2)}
\Big(1-e^{-ik\int_{\tau_1}^t(\lambda_1-\lambda_2)d\tau'}\Big)\Big]\, .
\end{eqnarray}
Problematic terms here are the exponentials
$\exp\Big[ik\int_{\tau_2}^{\tau_1}\lambda_jd\tau'\Big]$, where $j=1,2$.
Let us simplify these terms by introducing an average momentum function
\begin{equation}\label{bar_p}
\bar{p}_j=\int_{\tau_1}^{\tau_2}\frac{\lambda_j(\tau')d\tau'}{\tau_2-\tau_1}\, .
\end{equation}
Obviously, $\bar{p}_1=\bar{p}_2=\bar{p}(t)$, where we stressed that the averaged momentum function
is a function of time. Changing the integration from times
$(\tau_1,\tau_2)$ to $\tau=\tau_1$ and $s=\tau_1-\tau_2$, one recasts Eq. (\ref{action_clI})
in the form
\begin{eqnarray}\label{action_2}
&{}&\clI[\lambda_1(\tau),\lambda_2(\tau)]=-\frac{2\sigma^2}{\hh^2}\int_0^td\tau\int_{-t}^tds
\int_{-\infty}^{\infty}dk\int_{-\infty}^{\infty}d\omega
\nonumber \\
&\times& S(k,\omega) e^{i(k\bar{p}-\omega)s}
\Big[1-\cos\Big(k\int_{\tau}^t(\lambda_1-\lambda_2)d\tau'\Big)\Big]\, .
\end{eqnarray}
Integration over $s$ can be approximated by a $\delta$ function.  Namely, this integration yields
$$\int_{-t}^te^{i(k\bar{p}-\omega)s}ds\rightarrow
\int_{-\infty}^{\infty}e^{i(k\bar{p}-\omega)s}ds=2\pi\delta(\omega-k\bar{p})\, .$$
Now integration over the frequency $\omega$ can be performed that yields the action function
\begin{eqnarray}\label{action_3}
&{}&\clI[\lambda_1(\tau),\lambda_2(\tau)]=\clI[\lambda_1(\tau)-\lambda_2(\tau)]\nonumber \\
&=& -\frac{2\sigma^2}{\hh^2}\int_0^td\tau\int_{-\infty}^{\infty}dkS(k,k\bar{p})\nonumber \\
&\times & \Big[1-\cos\Big(k\int_{\tau}^t(\lambda_1-\lambda_2)d\tau'\Big)\Big]\, .
\end{eqnarray}
Finally,  one obtains
\[\clF[\lambda_1(\tau),\lambda_2(\tau)]=\exp\{\clI[\lambda_1(\tau)-\lambda_2(\tau)]\}\, ,\]
which is analogous to the expression obtained for Richardson diffusion. Performing again the variable change of Eq.
(\ref{lambda-mu-nu}), we obtain an expression for the MSD analogous to Eq. (\ref{MSD_1}). The MSD reads
\begin{eqnarray}\label{MSD_Hyper}
&{}&\langle x^2(t)\rangle=\int_{-\infty}^{\infty}\rho(x,t) x^2dx \nonumber \\
&=&\int\frac{dp_1dp_2}{4\pi^2}\bar{\Psi}_0(p_1)\bar{\Psi}_0^*(p_2)\delta^{(2)}(p_1-p_2)
\nonumber \\
&\times &\int\prod_{\tau}\frac{d\mu(\tau)d\nu(\tau)}{2\pi}
\exp\Big[i\int_0^t\mu(\tau)\nu(\tau)d\tau\Big] \nonumber \\
&\times & \exp\Big[-2i(p_1-p_2)\int_0^t\mu(\tau)d\tau \nonumber \\
&-&\frac{i\hh}{4}(p_1+p_2)\int_0^t\nu(\tau)d\tau\Big]
\clF[\nu(\tau)]\, .
\end{eqnarray}
The essential difference between Eqs. (\ref{MSD_Hyper}) and (\ref{MSD_1})
is the spectral density, which now is a two dimensional function $S(k,k\bar{p})$.
Integration over the fields $\mu$ and $\nu$ and differentiation over $p_1$ and $p_2$
yields
\begin{equation}\label{MSD_Hyper_2}
\langle x^2(t)\rangle=\pi\sigma^2
\int_0^td\tau(t-\tau)^2\int_{-\infty}^{\infty}S(k,k\bar{p})k^2dk
\end{equation}
We obtain the asymptotic behavior of Eq. (\ref{MSD_Hyper_2}) for large values of $\bar{p}$,
following a similar procedure presented in Refs.~\cite{fishman,wilkinson,fishman26}. Therefore,
by rescaling the variables, $k' = k\bar{p}$, one obtains
\begin{eqnarray}\label{MSD_Hyper_3}
\langle x^2(t)\rangle&=&\pi\sigma^2
\int_0^td\tau\frac{(t-\tau)^2}{\bar{p}^3(\tau)}
\int_{-\infty}^{\infty}S(\frac{k'}{\bar{p}},k')k'^2dk' \nonumber \\
&\approx & D_0\int_0^t\frac{(t-\tau)^2}{\bar{p}^3(\tau)}d\tau\, .
\end{eqnarray}
Here it was reasonable to suppose that $S(\frac{k}{\bar{p}},k)$ is a slow
function of $k/\bar{p}$.
For $\bar{p}={\rm const}$ the MSD corresponds to Richardson diffusion $\sim t^3$. Such behavior supposes for the averaged momentum function to be an increasing function of
time. Moreover, it has been suggested in Ref. \cite{fishman}
that for large $\bar{p}$, one obtains $S(k/\bar{p},k)\approx S(0,k)$ that yields nonzero
generalized diffusion coefficient
\begin{equation}\label{GDC}
D_0=\pi\sigma^2\int k^2S(0,k)dk\, .
\end{equation}
It is also supposes a physical meaning of $\bar{p}^2(t)$, which behaves as a
velocity-velocity correlation function. Therefore, one suggests a self-consistent procedure,
presented in Appendix A, to find this function. This yields for $\bar{p}(t)$
\begin{equation}\label{v-barp}
\bar{p}(t)=(5D_0/2)^{1/5}\, t^{1/5}\, .
\end{equation}
Taking this behavior into account, one obtains
\begin{equation}\label{main_result}
\langle x^2(t)\rangle\sim \bar{D} t^{12/5}\, ,
\end{equation}
which corresponds to hyperdiffusion, observed experimentally \cite{segev3}.
Here $\bar{D}=(2/5)^{3/5}D_0^{2/5}$

\section{Conclusion}

An enhanced spreading of a quantum wave packet in randomly inhomogeneous
media is considered. This quantum process is realized in an arbitrary space-time dependent potential $V(x,t)$. A rigorous quantum-mechanical calculation of the mean probability amplitude
(MPA) is suggested that makes it possible to calculate the mean
squared displacement (MSD) of the spreading wave packet. The obtained result establishes the power law spreading of the MSD, which is $\lgl x^2(t)\rgl\sim t^{\alpha}$, where  $2<\alpha\leq 3$, and the values of the transport exponent $\alpha$ depend on the correlation properties of the random potential $V(x,t)$.
The main motivation of the presented analysis is experimental demonstrations on wave packet spreading in random photonic lattices \cite{segev3,segev1,segev2}. Another possible
application of the presented analysis can be related to a sound waves monitoring in underwater acoustics \cite{UFN}, at the conditions when the parabolic equation approximation is valid
and the refractive index has random local fluctuations $\Delta n(x,z)$, which leads to a dominant random potential as in Eqs. (\ref{paraxial_eq}) and (\ref{Schrodinger}).

The rigorous formal expression for the wave function is obtained in a form of paths
integration, such that the wave function (\ref{sol_Psi}) is a functional of the random
potential $V(x,t)$. When $V(x,t)$ is $\delta$ correlated in time as in Eq. (\ref{CorrelatorC}),
the MSD is rigorously calculated in the framework of quantum-mechanical consideration.
The dominant term in the MSD  of the order of $t^3$ is  due to turbulent Richardson diffusion \cite{richardson}. Another important result of Eq (\ref{MSD_2}) is that the quantum homogeneous spread $\sim t^2$ stays separate from the dominant classical one $\sim t^3$. One can understand this property from the structure of the wave
function (\ref{sol_Psi})
\begin{eqnarray}\label{sol_Psi_a}
&{}&\Psi(x,t)=\int D[\lambda(\tau)]
\exp\left[\frac{i}{2\hh}\int_0^td\tau\lambda^2(\tau)\right]  \nonumber \\
&\times&e^{\frac{i}{\hh}\int_0^td\tau
V\Big(x+\int_{\tau}^t d\tau'\lambda\, ,\tau\Big)}
\Psi_0\Big(x+\int_0^td\tau\lambda\Big)\, , \nonumber \\
\end{eqnarray}
where $D[\lambda(\tau)]=\prod_{\tau}\frac{d\lambda(\tau)}{\sqrt{2\pi\hh i}}$.
This is a kind of Feynman-Kac formula \cite{feynman,kac}, obtained by means of the auxiliary
Markov process\footnote{Note that quantum mechanics itself is not the Markovian dynamics} with the Gaussian distribution in the potential $V(x,t)$. However, since $V(x,t)$ is random itself, the details of the potential are not important, and the main information, and contribution to the MPA is due to the correlation function $C(x-x',t-t')$, or the spectral density $S(k,\omega)$, correspondingly. When the random potential is $\delta$ correlated in time, the auxiliary field $\lambda$ does not intertwine with the potential $V(x,t)$.
This is reflected in the solution for the MPA $\rho(x,t)$, where the averaged
evolution kernel $\clF$ depends only on the quantum part of the auxiliary fields, namely $\clF=\clF[\lambda_1(\tau)-\lambda_2(\tau)]$. As a result of this, rigorous integration over $\lambda_1$ and $\lambda_2$ is performed. Therefore, each Markov process contributes separately
to the MSD in Eq. (\ref{MSD_2}).
The quantum mechanics leads to the ballistic $\sim t^2$
spread of the initial wave packet, while the classical Obukhov mechanism of
turbulent diffusion reveals itself in pure quantum mechanics with the dominant $\sim t^3$ spread
of the wave packet.

The situation changes dramatically, when the random potential is correlated in both space and time. In this case the auxiliary $\lambda$ fields and the random potential are intertwined due to the nonlocal terms $\int_{t_1}^{t_2}\lambda(\tau)d\tau$ in the MPA. To make the problem treatable, this nonlocal term is presented in form of an averaged quantum momentum function $(t_1-t_2)\bar{p}$,
where $\bar{p}$ is related to a velocity-velocity correlation function of random quantum paths.
After this approximation, the integration
over the $\lambda$s is performed rigorously again. Now the quantum ballistic spread is accompanied by hyperdiffusion $\sim t^{\alpha}$. Assuming  that the spectral function after rescaling
$S(\frac{k}{\bar{p}},k)$ is a slow function of $k/\bar{p}$, like in Eq. (\ref{GDC}), it is obtained that $\alpha=12/5$.
As already admitted, this result coincides with one obtained in Refs.~\cite{wilkinson,fishman}
in the classical limit of the ray dynamics. However, contrary to Refs.~\cite{wilkinson,fishman},
in the present analysis we did not suppose any restriction conditions for
the random potential $V(x,t)$.

In the general case, one obtains that $2<\alpha<3$. This result follows from Eq.
(\ref{MSD_Hyper_3}), where
$\int_{-\infty}^{\infty}S(\frac{k}{\bar{p}},k)k^2dk$ is a slow varying function, which approaches to the transport constant $D_0$ for the asymptotic large times $t\rightarrow\infty$.
Therefore, $\alpha=12/5$ is the large time asymptotic result, as well.

\section*{Acknowledgments}

I thank Professor S. Fishman for helpful and informative discussions,
and comments to the text. This research was supported by the Israel
Science Foundation (ISF-1028).

\appendix

\section{Inferring of the averaged momentum $\bar{p}$}

Let us obtain analytical expression (\ref{v-barp}) for the averaged momentum $\bar{p}(t)$
in the framework of a self-contained procedure, where we take into account that
$\bar{p}^2(t)$ is a correlation function.
First, it is worth noting that Eq. (\ref{bar_p}) is a definition of $\bar{p}$. However
it does not determine the latter, since the averaging of the random auxiliary field $\lambda(t)$ over the time interval $s=t_1-t_2$ is not well defined.
Second, we admit that the integral $\int_{\tau_2}^{\tau_1}\lambda(t)dt$ is not zero.
Moreover, we replace this integral by a quantum path. One can reasonably suppose
that the MSD of these quantum paths is determined by the velocity-velocity correlation
function
$\bar{p}^2(t)\approx\lgl\dot{x}^2(t)\rgl_{\lambda}=\lgl\dot{x}^2(t)\rgl$. Therefore, we relate the averaged momentum function to the real quantum path $x(t)$, and to estimate its temporal behavior, we consider its classical random dynamics.
Integrating the dynamical equation $$\ddot{x}=-\frac{d\,V(x,t)}{d\,x}\, , $$
one obtains from the definition of the random potential in Eq. (\ref{potentialV})
\begin{equation}\label{A1}
\dot{x}(t)=\int_0^t\frac{1}{\sqrt{N}}\sum_m(-i)k_mA_me^{ik_mx(t')-i\omega_mt'}dt'+c.c.\, .
\end{equation}
Therefore, the self-correlation function reads
\begin{eqnarray}\label{A2}
\lgl\dot{x}^2(t)\rgl&=&\frac{2\sigma^2}{N}\sum_m\int_0^tdt'\int_0^tdt^{''} \nonumber  \\
&\times& \cos[k_m(x'-x'')-\omega(t'-t'')]k_m^2 \nonumber \\
&=& 2\sigma^2\int_0^tdt'\int_0^tdt^{''}\int dk\int d\omega k^2S(k,\omega) \nonumber \\
&\times&  \cos[k(x'-x'')-\omega(t'-t'')]\,.
\end{eqnarray}
Note that according the property (\ref{EAm}), $\lgl\dot{x}(t)\rgl=0$.
After changing integration times $\tau=t'$ and $s=t'-t''$,
we arrived at the same expression as in Eq. (\ref{action_2}).

Following the solution of the wave function in Eq. (\ref{sol_Psi}),
the evolution of the coordinates $x(t)$ is due to the shift operator $x(t)=x_0+\int_0^t\lambda(\tau)d\tau$.
Therefore, according Eq. (\ref{bar_p}), the difference $x(t')-x(t'')$ yields
$\bar{p}(\tau)=(x'-x'')/(t'-t'')=\dot{x}$.
Taking into account Eq. (\ref{MSD_Hyper_3}), one obtains approximately from Eq. (\ref{A2})
\begin{equation}\label{A3}
\bar{p}^2(t)\approx D_0\int_0^t\frac{d\tau}{\bar{p}^3(\tau)}\, .
\end{equation}
Differentiating Eq. (\ref{A3}) over time, one obtains
\begin{equation}\label{A4}
\bar{p}^3(t)\frac{d\,\bar{p}^2(t)}{d\,t}=D_0\,.
\end{equation}
This equation can be also obtained by using the well known expression
(see \textit{e.g.}, \cite{Reif})
\begin{equation}\label{approx_p}
\lgl x^2(t)\rgl_{\lambda}=2\int_0^t(t-\tau)\lgl\dot{x}(\tau)\dot{x}(0)\rgl_{\lambda}d\tau
\end{equation}
and consider that $\lgl x^2(t)\rgl= \lgl x^2(t)\rgl_{\lambda}$.
Now we put forward the physical meaning of the momentum function $\bar{p}$ by substituting
it in Eq. (\ref{approx_p})
\begin{equation}\label{approx_p2}
\lgl x^2(t)\rgl_{\lambda}=2\int_0^t(t-\tau)\bar{p}^2(\tau)d\tau\, .
\end{equation}
Differentiating twice Eqs. (\ref{approx_p2}) and (\ref{MSD_Hyper_3}) over $t$ and
comparing the obtained results, one obtains
\[\bar{p}^2\approx D_0\int_0^t\frac{d\tau}{\bar{p}^3(\tau)}\, .\]
Differentiating this over time again, one obtains Eq. (\ref{A4}). Solving this equation,
one obtains
\begin{equation}\label{sol_bar_p}
\bar{p}(t)=(5D_0/2)^{1/5}t^{1/5}\, .
\end{equation}


\begin{thebibliography}{99}

\bibitem{segev3} L. Levi, Y. Krivolapov, S. Fishman and M. Segev,
Nature Phys. \textbf{8}, 912 (2012).

\bibitem{wilkinson} E. Arvedson, M. Wilkinson, B. Mehlig, and K. Nakamura, Phys. Rev. Lett.
\textbf{96}, 030601 (2006).

\bibitem{fishman} Y. Krivolapov, L. Levi, S. Fishman, M. Segev and M. Wilkinson,
New J. Phys. \textbf{14}, 043047 (2012).

\bibitem{segev1}) L. Levi, M. Rechtsman, B. Freedman,
T. Schwartz, O. Manela and M. Segev,
Science \textbf{332}, 1541 (2011).

\bibitem{segev2} M. Rechtsman, L. Levi, B. Freedman, T. Schwartz,
O. Manela and M. Segev,
Optics and Photonics News (Special Issue: Optics in 2011) \textbf{22} (12), (2011).

\bibitem{kliatskin} V.I. Kliatskin, \textit{Stochastic equations and waves in
randomly inhomogeneous media} (Nauka, Moscow, 1980) (in Russian).

\bibitem{dashen} R. Dashen J. Math. Phys. \textbf{20} 894 (1979); Opt. Lett. \textbf{9},
110 (1984).

\bibitem{zj} J. Zinn-Justin, \textit{Quantum Field Theory and Critical Phenomena}
(Claredon Press, Oxford, 1990).

\bibitem{UPN2004} V.I. Kliatskin, Phys. Uspekhi \textbf{47}, 169  (2004).

\bibitem{tappert} E. D. Tappert, The Parabolic Approximation Method, Lectures Notes in Physics,
\textbf{70}, in: Wave Propagation and Underwater Acoustics, eds. by J. B. Keller and
J. S. Papadakis, (Springer, New York, 224-287, 1977).

\bibitem{richardson} L. F. Richardson, Proc. R. Soc. A \textbf{110}, 709 (1926).

\bibitem{frisch} U. Frisch, \textit{Turbulence. The Legacy of Kolmogorov} (Cambridge Univ.
Press, Cambridge, 1995).

\bibitem{monin_yaglom} A.S. Monin, A.M. Yaglom, \textit{Statistical Fluid Mechanics:
Mechanics of Turbulence} vol. 1 ( MIT Press, Cambridge, 1971); \textit{ibid} vol 2 
(MIT Press, Cambridge, 1975).

\bibitem{obukhov} A.M. Obukhov, Adv. Geophys. \textbf{6}, 113 (1959).

\bibitem{prl48} A.M. Jayannavar and N. Kumar, Phys. Rev. Lett. \textbf{48},
553 (1982).

\bibitem{furutsu}  K. Furutsu, J. Res. N.B.S. D\textbf{67}, 303 (1963).

\bibitem{novikov}
E.A. Novikov Zh. Eksp. Teor. Fiz. \textbf{47}, 1919 (1964) [Sov. Phys. JETP
\textbf{20}, 1990 (1965)].

\bibitem{barkai} D.A. Kessler and E. Barkai, Phys. Rev. Lett. \textbf{108}, 230602 (2012).

\bibitem{nir} Y. Sagi, M. Brook, I. Almog, and N. Davidson, Phys. Rev. Lett.
\textbf{108}, 093002 (2012).

\bibitem{feynman} R.P. Feynman and A.R. Hibbs, \textit{Quantum Mechanics and Path
Integrals} (McGrow Hill, New York, 1965).

\bibitem{kac} M. Kac, \textit{Probability and Related Topics in Physical
Sciences}. (Interscience, New York, 1958).

\bibitem{UFN} A.L. Virovlyansky, D.V. Makarov, and S.V. Prants, Phys. Uspekhi
\textbf{55}, 18 (2012).

\bibitem{tomsovic}
K. C. Hegewisch and S. Tomsovic  Europhys. Lett. \textbf{97}, 34002 (2012).

\bibitem{leontovich} M.A. Leontovich, Izv. USSR Ac.Sc., Phys. \textbf{8}, 16, 1944 (in Russian).

\bibitem{khohlov} R.V. Khohlov,  Radiotech. and Elrctron. \textbf{6}, 1116, 1961 (in Russian).

\bibitem{Strat} R.L. Stratonovich, Dokl. Akad. Nauk SSSR \textbf{115}, 1097 (1957)
[Sov. Phys. Dokl. \textbf{2}, 416 (1958)].

\bibitem{Hab} J. Hubbard, Phys. Rev. Lett. \textbf{3}, 77 (1958).

\bibitem{sokolov} V.V. Sokolov, Theor. Math. Phys. \textbf{61}, 1041 (1984).

\bibitem{bi1987} G.P. Berman and A.M. Iomin, Theor. Math. Phys. \textbf{77}, 1197 (1988).

\bibitem{pre70_2004} A. Iomin, Phys. Rev. E \textbf{70}, 026206 (2004).

\bibitem{karruthers_nieto} P. Carruthers and M.M. Nieto, Rev. Mod. Phys. \textbf{40},
411 (1968).

\bibitem{falkovich} G. Falkovich, K. Gawedzki, and M. Vergassola,
Rev. Mod. Phys. \textbf{73}, 913 (2001).

\bibitem{baule} A. Baule and R. Friedrich, Phys. Lett. A \textbf{350}, 167 (2006).

\bibitem{fishman26}  V. Bezuglyy, B. Mehlig, M. Wilkinson, K. Nakamura, and E. Arvedson,
J. Math. Phys. \textbf{47}, 073301 (2006).

\bibitem{Reif} F. Reif, \textit{Fundamentals of statistical and thermal physics}
(McGraw Hills, New York, 1965).


\end{thebibliography}
\end{document}